\documentclass[a4paper,11pt]{article}
\usepackage{pos}
\usepackage{amsmath,bm,longtable,mathrsfs,slashed}

\renewcommand{\logo}{\relax}

\title{Axial U(1) symmetry near the pseudocritical temperature in $N_f=2+1$ lattice QCD with chiral fermions}
\ShortTitle{Axial U(1) symmetry near the pseudocritical temperature in $N_f=2+1$ lattice QCD with chiral fermions}

\manuallySeparateAuthors
\author[a,b]{Sinya Aoki}
\author[c]{, Yasumichi Aoki}
\author[d]{, Hidenori Fukaya}
\author[e,f]{, Shoji Hashimoto}
\author[c]{, \\Issaku Kanamori}
\author[e,f,g]{, Takashi Kaneko}
\author[c]{, Yoshifumi Nakamura}
\author[d]{, \\Christian Rohrhofer}
\author[h]{, Kei Suzuki}
\author[]{, and}
\author[d]{ David Ward}
\author[]{ (JLQCD Collaboration)}

\affiliation[a]{Center for Gravitational Physics, Yukawa Institute for Theoretical Physics, Kyoto 606-8502, Japan}
\affiliation[b]{RIKEN Nishina Center (RNC), Saitama 351-0198, Japan}
\affiliation[c]{RIKEN Center for Computational Science, Kobe 650-0047, Japan}
\affiliation[d]{Department of Physics, Osaka University, Toyonaka 560-0043, Japan}
\affiliation[e]{KEK Theory Center, High Energy Accelerator Research Organization (KEK), Tsukuba 305-0801, Japan}
\affiliation[f]{School of High Energy Accelerator Science, The Graduate University for Advanced Studies (Sokendai), Tsukuba 305-0801, Japan}
\affiliation[g]{Kobayashi-Maskawa Institute for the Origin of Particles and the Universe, Nagoya University, Nagoya 464-8602, Japan}
\affiliation[h]{Advanced Science Research Center, Japan Atomic Energy Agency (JAEA), Tokai 319-1195, Japan}

\emailAdd{k.suzuki.2010@th.phys.titech.ac.jp}

\abstract{
We study the $U(1)_A$ anomaly at high temperatures of $N_f=2+1$ lattice QCD
with chiral fermions. Gauge ensembles are generated with M\"obius
domain-wall (MDW) fermions, and the measurements are reweighted to those with overlap fermions.
We report on the results for the Dirac spectra, the $U(1)_A$ susceptibility, and the topological
susceptibility in the temperature range of $T=136$, $153$, $175$, and $204$ MeV, where the up and down quark masses are set to be near the physical point as well as at lighter or heavier masses.
}

\FullConference{%
 The 40th International Symposium on Lattice Field Theory, LATTICE2023\\
  July 31st - August 4th, 2023\\
  Fermi National Accelerator Laboratory
}

\begin{document}
\begin{flushright}
KEK-CP-0399, OU-HET-1216
\end{flushright}
\maketitle

\section{Introduction}\label{sec-1}

In quantum chromodynamics (QCD) describing dynamics of the quarks and gluons, key features are the $SU(2)_L \times SU(2)_R$ chiral symmetry (for up and down quarks near the massless limit) and the $U(1)_A$ symmetry.
In the low-temperature phase of QCD, the chiral symmetry is spontaneously broken due to the chiral condensate, while the $U(1)_A$ symmetry is explicitly broken by the quantum anomaly.
In the high-temperature phase, the chiral condensate is suppressed and the chiral symmetry is restored (for massless quarks), while the temperature-dependence of the $U(1)_A$ anomaly is not undestood well (for recent progress in lattice QCD simulations, see Ref.~\cite{Bazavov:2019www,Ding:2020xlj,Kaczmarek:2021ser,Kaczmarek:2023bxb}).

The JLQCD collaboration has been working on this issue~\cite{Cossu:2013uua,Tomiya:2016jwr,Aoki:2020noz,Aoki:2021qws} by using lattice QCD simulations with chiral fermions, i.e., the overlap (OV) fermion formulation~\cite{Neuberger:1997fp,Neuberger:1998my}.
Since the OV fermion formulation keeps tha chiral symmetry on the lattice and should be appropriate to examine physics related to the chiral and $U(1)_A$ symmetries.
However, simulations with OV fermions require an enormous computational cost, so that we have to apply a technique to reduce the cost.
For example, in Ref.~\cite{Cossu:2013uua} the topological sector was fixed to the trivial one.
In Ref.~\cite{Tomiya:2016jwr,Aoki:2020noz,Aoki:2021qws}, the MDW/OV reweighting  technique \cite{Fukaya:2013vka,Tomiya:2016jwr} was applied, where physical quantities measured on gauge ensembles generated with M\"obius domain-wall (MDW) fermions~\cite{Brower:2004xi,Brower:2005qw,Brower:2012vk} are reweighted to corresponding ones on OV fermion ensembles. 
So far, our simulations were mainly done with two flavors of quarks.

In this work, we study $N_f=2+1$ QCD including the strange quark.
In this contribution, we report on the preliminary results from $N_f=2+1$ lattice QCD simulations.
Some of results in high-temperature phase ($T=204$ MeV and $175$ MeV) were already reported in our previous proceedings~\cite{Aoki:2022ebi}, and here main updates are the results at lower temperatures: $T=153$ MeV near the pseudocritical temperature and $136$ MeV in the chiral symmtry broken phase.\footnote{If the pseudocritical temperature is assumed to be $T_c = 155$ MeV, the four temperatures $T=204$, $175$, $153$, and $136$ MeV are approximately $T \sim 1.3T_c$, $1.1T_c$, $1.0 T_c$, and $0.9T_c$, respectively.}
Parameters of numerical lattice simulations are summarized in Table \ref{Tab:param}.
We applied the tree-level Symanzik improved gauge action~\cite{Luscher:1985zq} and dynamical MDW fermion action~\cite{Brower:2004xi,Brower:2005qw,Brower:2012vk}, and measured quantities are reweighted to those with overlap fermions.
The gauge coupling is set to be $\beta=4.17$ which corresponds to the lattice cutoff $a^{-1}=2.453$ GeV (or the lattice spacing $a\sim 0.08$ fm) determined in our simulations at zero temperature~\cite{Colquhoun:2022atw}.
We generated gauge ensembles with the up and down quark masses in the range of $am_{ud}=0.002$-$0.0120$, where $am_{ud}=0.002$ ($m_{ud} \sim 4.9$ MeV) is considered to be near the physical point (if $m_{ud}^\mathrm{phys} \sim 3.5$ MeV).
In addition, the observables at $am_{ud}=0.001$ ($m_{ud} \sim 2.5$ MeV lighter than the physical point) are obtained by the mass reweighting technique from the ensembles at $am_{ud}=0.002$.
The strange-quark mass is fixed to be $am_s=0.04$ ($m_s \sim 98.1$ MeV near the physical point).

\begin{table}[t!]
\centering
\caption{Numerical parameters of lattice simulations.
$L^3 \times L_t $ are lattice size in the spatial and temporal directions.
$m_{ud}$ and $m_s$ are the degenerate mass of up and down quarks and strange-quark mass, respectively.
}
\small
\begin{tabular}{cccccc}
\hline\hline
$L^3 \times L_t $ &  $T$ (MeV) & $am_{ud}$ & $m_{ud}$ (MeV) & $am_s$ & Comments \\
\hline
$32^3 \times 12$ & 204 & 0.0010  & 2.5 & 0.040 & $m_{ud}$ by mass reweighting\\
$32^3 \times 12$ & 204 & 0.0020  & 4.9 & 0.040 &\\
$32^3 \times 12$ & 204 & 0.0035  & 8.6 & 0.040 &\\
$32^3 \times 12$ & 204 & 0.0070  & 17 & 0.040 &\\
$32^3 \times 12$ & 204 & 0.0120  & 29 & 0.040 &\\
\hline
$32^3 \times 14$ & 175 & 0.0010  & 2.5 & 0.040 & $m_{ud}$ by mass reweighting\\
$32^3 \times 14$ & 175 & 0.0020  & 4.9 & 0.040 &\\
$32^3 \times 14$ & 175 & 0.0035  & 8.6 & 0.040 &\\
$32^3 \times 14$ & 175 & 0.0050  & 12  & 0.040 &\\
$32^3 \times 14$ & 175 & 0.0070  & 17  & 0.040 &\\
$32^3 \times 14$ & 175 & 0.0120  & 29  & 0.040 &\\
\hline
$32^3 \times 16$, $40^3 \times 16$ & 153 & 0.0010  & 2.5 & 0.040 & $m_{ud}$ by mass reweighting\\
$32^3 \times 16$, $40^3 \times 16$ & 153 & 0.0020  & 4.9 & 0.040 &\\
$32^3 \times 16$, $40^3 \times 16$ & 153 & 0.0035  & 8.6 & 0.040 &\\
$32^3 \times 16$, $40^3 \times 16$ & 153 & 0.0070  & 17  & 0.040 &\\
$32^3 \times 16$, $40^3 \times 16$ & 153 & 0.0120  & 29  & 0.040 &\\
\hline
$36^3 \times 18$, $48^3 \times 18$ & 136 & 0.0010  & 2.5 & 0.040 & $m_{ud}$ by mass reweighting\\
$36^3 \times 18$, $48^3 \times 18$ & 136 & 0.0020  & 4.9 & 0.040 &\\
$36^3 \times 18$, $48^3 \times 18$ & 136 & 0.0035  & 8.6 & 0.040 &\\
$36^3 \times 18$, $48^3 \times 18$ & 136 & 0.0070  & 17  & 0.040 &\\
$36^3 \times 18$, $48^3 \times 18$ & 136 & 0.0120  & 29  & 0.040 &\\
\hline\hline
\end{tabular}
\label{Tab:param}
\end{table}

\section{Overlap Dirac spectrum}\label{sec-2}
First, we investigate the eigenvalue spectrum of the overlap Dirac operator,
\begin{equation}
\rho(\lambda)=\frac{1}{V} \sum_{\lambda_i} \langle \delta(\lambda-\lambda_i) \rangle,
\end{equation}
where $V$ is the volume, and $\lambda_i$ is the $i$-th positive
eigenvalue of the overlap Dirac operator.
$\langle \mathcal{O} \rangle$ means the gauge ensemble average of a quantity $\mathcal{O}$.
As is well known, the Banks-Casher relation~\cite{Banks:1979yr}, $\langle \bar{q} q \rangle = - \lim_{V\to \infty} \pi \rho(0)$, states that the chiral condensate $\langle \bar{q} q \rangle$ is proportional to the value of Dirac spectrum at $\lambda=0$.

\begin{figure}[t!]
    \begin{minipage}[t]{0.5\columnwidth}
    \includegraphics[clip,width=1.0\columnwidth]{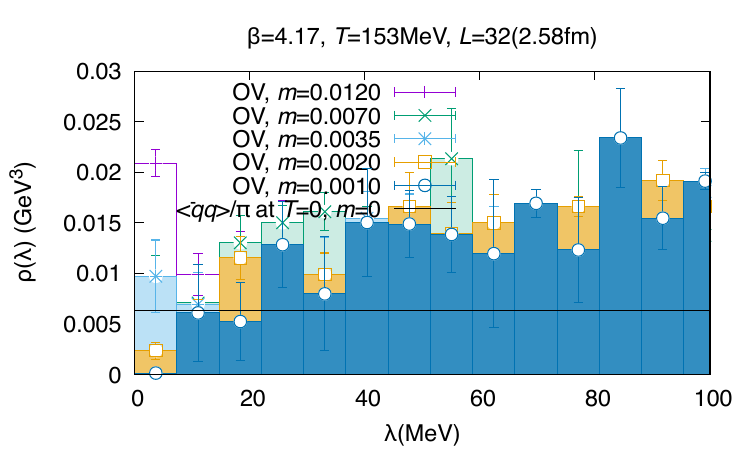}
    \end{minipage}%
    \begin{minipage}[t]{0.5\columnwidth}
    \includegraphics[clip,width=1.0\columnwidth]{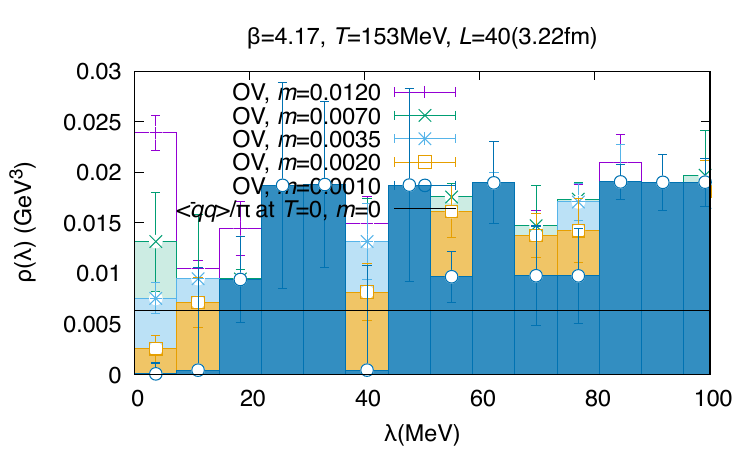}
    \end{minipage}
    \begin{minipage}[t]{0.5\columnwidth}
    \includegraphics[clip,width=1.0\columnwidth]{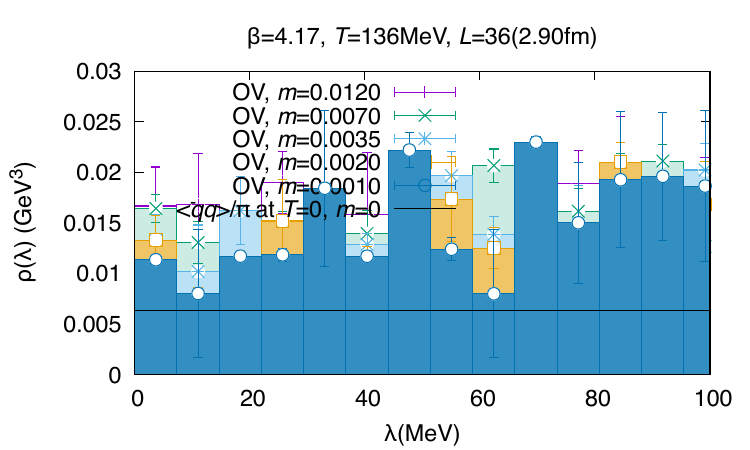}
    \end{minipage}%
    \begin{minipage}[t]{0.5\columnwidth}
    \includegraphics[clip,width=1.0\columnwidth]{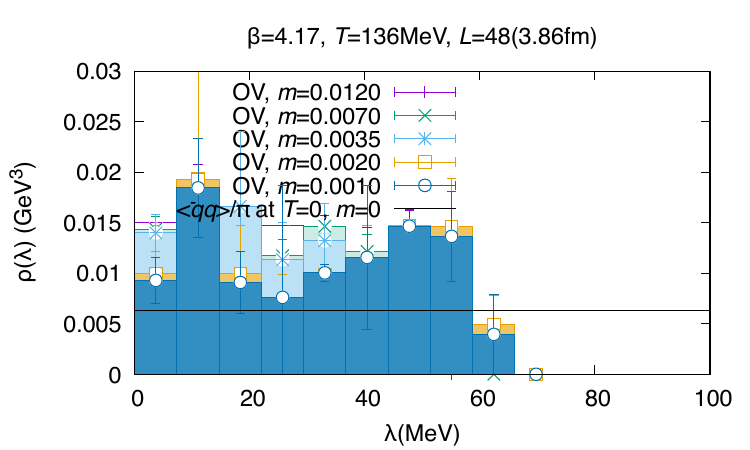}
    \end{minipage}
    \caption{Dirac spectrum $\rho(\lambda)$ on reweighted OV ensembles at $T=153$ (upper) and $136$ MeV (lower).
Results with two volumes are shown in left and right panels.
}
    \label{fig:spectrum}
\end{figure}

In Fig.~\ref{fig:spectrum}, we show the Dirac spectra at $T=153$ MeV (top panels) and 136 MeV (bottom panels).
$T=153$ MeV is near (or slightly lower than) the pseudocritical temperature in $N_f=2+1$ QCD with the physical quark mass.
While $\rho(0)$ is nonzero for heavier quark mass ($m_{ud} \gtrsim 4.9$ MeV), it is consistent with zero at the lightest quark mass ($m_{ud} \sim 2.5$ MeV), which suggests $\rho(0) \sim 0$ in the chiral limit ($m_{ud} \to 0$).
On the other hand, since $T=136$ MeV is below the pseudocritical temperature, $\rho(0)$ is nonzero at all the quark masses.
Note that the horizontal line is the chiral condensate at zero temperature (divided by $\pi$), which was obtained in our previous simulations \cite{Cossu:2016eqs}.
For Dirac spectra at higher temperatures ($T=204$ and 175 MeV), see our previous proceedings~\cite{Aoki:2022ebi}.

\begin{figure}[t!]
    \begin{center}
            \includegraphics[clip,width=0.5\columnwidth]{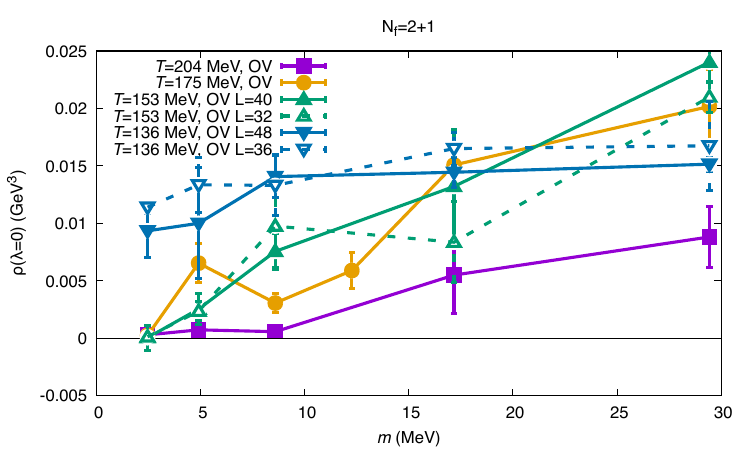}
    \end{center}
    \caption{Quark mass dependence of lowest-bin value of Dirac spectrum, $\rho(\lambda=0)$, at $T=136$-$204$ MeV.
}
    \label{fig:lowest_bin}
\end{figure}

In Fig.~\ref{fig:lowest_bin}, we show the value of the the lowest bin of the Dirac specturm as a function of $m_{ud}$.
We confirm that, at the lightest quark mass, the lowest bin is strongly suppressed at $T=153$-$204$ MeV, which is suggests that it is almost zero even in the chiral limit. 
On the other hand, at $T=136$ MeV, the value of the lowest bin is nonzero, which is due to the chiral condensate.

\section{$U(1)_A$ susceptibility}\label{sec-3}

The $U(1)_A$ susceptibility is defined as the
difference between the two-point mesonic correlators of the isovector-pseudoscalar channel $\pi^a$ and
isovector-scalar channel $\delta^a$ (the subscript $a$ is the isospin index):
\begin{equation}
\Delta_{\pi-\delta} \equiv \chi_\pi - \chi_\delta \equiv \int d^4x \langle \pi^a(x) \pi^a(0) - \delta^a (x) \delta^a(0) \rangle. \label{eq:Delta_def}
\end{equation}
On the lattice, $\Delta_{\pi-\delta}$ is calculated by the spectral decomposition of the overlap Dirac operator~\cite{Cossu:2015kfa},
\begin{equation}
\Delta_{\pi-\delta}^{\mathrm{ov}} =  \frac{1}{V(1-m^2)^2} \left< \sum_i \frac{2m^2(1-\lambda_i^{(\mathrm{ov},m)2})^2}{\lambda_i^{(\mathrm{ov},m)4}} \right>, \label{eq:Delta_ov}
\end{equation}
where $\lambda_i^{(\mathrm{ov},m)}$ is the eigenvalue of the massive overlap Dirac operator $H_m=\gamma_5 [(1-m)D_\mathrm{ov}+m]$ with a mass $m$, and the lattice spacing is set as $a=1$.\footnote{Note that this is one of the simplest definitions with overlap Dirac eigenvalues.
One can define other $U(1)_A$ susceptibilities by subtracting the contributions from chiral zero modes~\cite{Tomiya:2016jwr,Aoki:2020noz} and/or from the logarithmic UV divergence~\cite{Aoki:2020noz}.
In this work, we show results obtained without such subtraction schemes.}

\begin{figure}[t!]
    \begin{center}
           \includegraphics[clip,width=0.5\columnwidth]{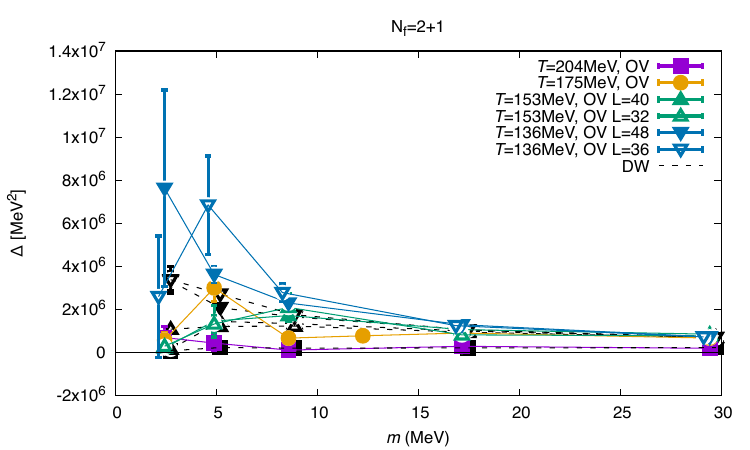}
    \end{center}
    \caption{Quark mass dependence of $U(1)_A$ susceptibility $\Delta_{\pi-\delta}^{\mathrm{ov}}$ at $T=136$-$204$ MeV.
}
    \label{fig:u1_sus}
\end{figure}

In Fig.~\ref{fig:u1_sus}, we show the results for the $U(1)_A$ susceptibility.
At $T=153$-$204$ MeV, we find that the $U(1)_A$ susceptibility is strongly suppressed at the lightest quark mass, which suggests that the $U_1(A)$ susceptibility is consistent with zero in the chiral limit.
On the other hand, in the chiral symmetry broken phase ($T=136$ MeV), it is nonzero.

\section{Topological susceptibility}\label{sec-5}

The topological susceptibility is defined as
\begin{equation}
\chi_t \equiv \frac{\langle Q_t^2 \rangle}{V}, \label{chit}
\end{equation}
with the gauge ensemble average of the topological charge $Q_t$ for the gluon fields.
In order to compute the topological charge, we apply two types of methods:
(i) In the first method, the topological charge is obtained from the index of the overlap Dirac operator,
\begin{equation}
Q_t=n_+ - n_-,  \label{Q_fermion}
\end{equation} 
where $n_\pm$ is the number of the zero modes with positive ($+$) or negative (-) chirality.
(ii) In another method, the topological charge is defined by the field-strength tensor of gluon fields.
In order to define it on the lattice, we apply the gradient flow \cite{Luscher:2010iy} in the space of lattice gauge fields.
Then,
\begin{equation}
Q_t (t)=\frac{1}{32\pi^2} \sum_x \varepsilon^{\mu \nu \rho \sigma} \mathrm{Tr} \, \left[ F_{\mu \nu}(x,t) F_{\rho \sigma} (x,t) \right], \label{Q_gluon}
\end{equation}
where $F_{\mu \nu}(x,t)$ is a field-strength tensor of lattice gauge fields at spacetime $x$ and a flow time $t$.
In this work, we adopt the clover-type construction of $F_{\mu \nu}(x,t)$, and we pick the results at $ta^2=5$ (a plateau for $Q_t (t)$ is stable for $ta^2 \gtrsim 3$~\cite{Bruno:2014ova}).

\begin{figure}[t!]
    \begin{minipage}[t]{0.5\columnwidth}
    \includegraphics[clip,width=1.0\columnwidth]{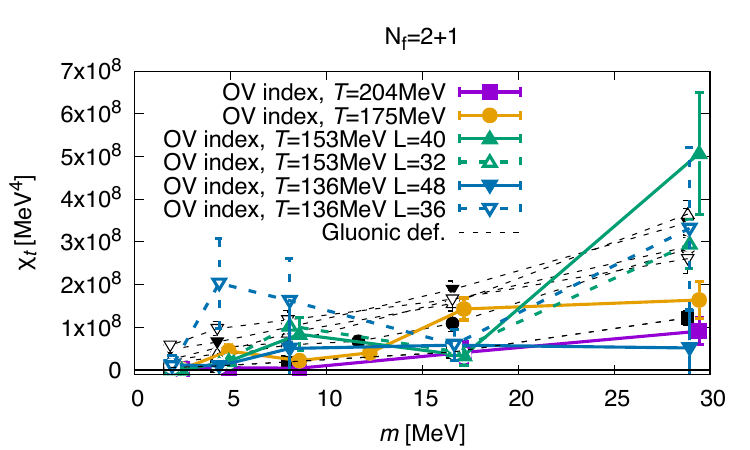}
    \end{minipage}%
    \begin{minipage}[t]{0.5\columnwidth}
    \includegraphics[clip,width=1.0\columnwidth]{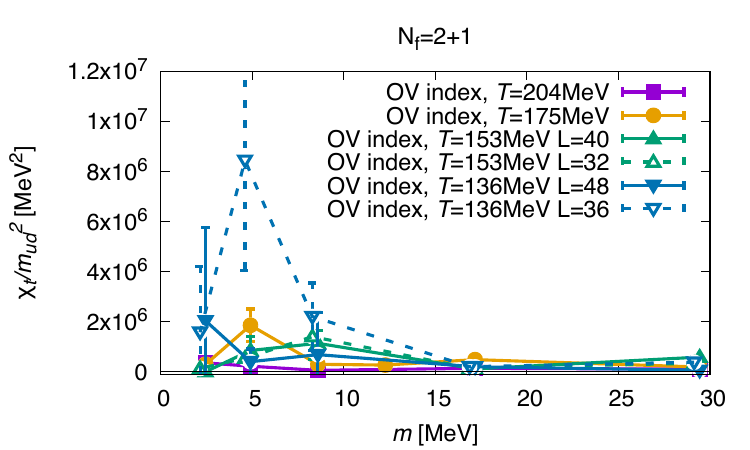}
    \end{minipage}
    \caption{Quark mass dependence of topological susceptibilities $\chi_t$ and its divided quantities $\chi_t/m_{ud}^2$ at $T=136$-$204$ MeV.
In the left panel, colored and black points represent $\chi_t$ from the fermionic definition (\ref{Q_fermion}) on reweighted OV ensembles and $\chi_t$ from the gluonic definition (\ref{Q_gluon}) on MDW ensembles, respectively.
}
    \label{fig:chit}
\end{figure}

In Fig.~\ref{fig:chit}, we show the results for the quark mass dependence of the topological susceptibility.
Since $\chi_t$ is expected to be proportional to $m_{ud}$ in the chiral symmetry broken phase (e.g., at the leading order of the chiral perturbation theory~\cite{Leutwyler:1992yt}), we plot another quantity $\chi_t/m_{ud}^2$ in order to check the scaling with $m_{ud}$.
From the right panel, we find that $\chi_t/m_{ud}^2$ at $T=153$-$204$ MeV is strongly suppressed toward the chiral limit, whereas the result at $T=136$ MeV stays at nonzero values.
Thus, the behaviors of $\chi_t$ in the chiral symmetry broken phase and in the high-temperature phase are distinct.

\section{Conclusion and outlook}\label{sec-6}

In this contribution, we show the preliminary results for (i) Dirac spectra $\rho(\lambda)$, (ii) $U(1)_A$ suceptibilities $\Delta_{\pi-\delta}$, and (iii) topological suscptibilities $\chi_t$ in $N_f=2+1$ lattice QCD simulations with chiral (i.e., reweighted overlap) fermions
in the temperature range of $T=136$-$204$ MeV.
Dependences of these quantities on up and down quark masses are studied in the range of $m_{ud}=2.5$-$29$ MeV, which covers not only the physical quark mass but also a lighter quark mass (i.e., near the chiral limit).
Our results suggest that $\rho(0)$, $\Delta_{\pi-\delta}$, and $\chi_t/m_{ud}^2$ in the chiral limit at $T=153$-$204$ MeV (near and higher than the pseudocritical temperature) is strongly suppressed, while those at $T=136$ MeV (in the low-temperature phase) are nonzero. 

Finally, we give remarks on other measurements:
\begin{itemize}
\item {\it Chiral susceptibility}---The chiral susceptibility~\cite{Karsch:1994hm}, defined as the first derivative of the chiral condensate with respect to the quark mass, is useful for estimating the pseudocritical temperature for the chiral condensation.
Also, it can be related to the $U(1)_A$ and topological suscptibilities as well as the chiral condensate.
Our previous results~\cite{Aoki:2021qws} at $N_f=2$ suggest that both the connected and disconnected parts of the chiral susceptibility are dominated by the contribution from the $U(1)_A$ anomaly in the region of $m_{ud} \ge 2.64$ MeV at $T=220$-$330$ MeV and also in $m_{ud} \ge 6.6$ MeV at $T=190$ MeV (see Ref.~\cite{Aoki:2021vji} for $T=165$ MeV).
For our latest updates, see Ref.~\cite{Aoki:2024uvl}.
\item {\it Hadronic correlators}---Also from mesonic/baryonic correlators, we can study the chiral and $U(1)_A$ symmetries from the viewpoint of a degeneracy between the correlators for their symmetry partners.
In addition, in the high-temperature phase, not only the chiral and $U(1)_A$ symmetries but also an emergent symmetry can appear, which is called the chiral spin symmetry~\cite{Glozman:2014mka, Glozman:2015qva} (see Ref.~\cite{Glozman:2022zpy} for a recent review).
The results~\cite{Rohrhofer:2017grg,Rohrhofer:2019qwq} for mesonic spatial correlators in $N_f=2$ lattice QCD suggest that this symmetry is realized in the range of $T=220$-$480$ MeV (see Ref.~\cite{Rohrhofer:2019qal} for the temporal correlators).
For our latest updates including $N_f=2+1$ simulations, see Ref.~\cite{Ward:2024tdm}.
\end{itemize}

\section*{Acknowledgment}\label{sec-Ack}
We thank Y. Sumino for useful discussion.
We used the QCD software packages Iroiro++~\cite{Cossu:2013ola}, Grid~\cite{Boyle:2015tjk,Meyer:2021uoj}, and Bridge++~\cite{Ueda:2014rya,Akahoshi:2021gvk}.
Numerical simulations are performed on IBM System Blue Gene Solution at KEK under a support of its Large Scale Simulation Program (No. 16/17-14), Oakforest-PACS at JCAHPC under a support of the HPCI System Research Projects (Project IDs: hp170061, hp180061, hp190090, hp200086, and hp210104) and Multidisciplinary Cooperative Research Program in CCS, University of Tsukuba (xg17i032 and xg18i023), the supercomputer Fugaku provided by the RIKEN Center for Computational Science (hp200130, hp210165, hp210231, hp220279, and hp230323), Wisteria/BDEC-01 Odyssey at JCAHPC (HPCI: hp220093 and hp230070, MCRP: wo22i038), and Polarie and Grand Chariot at Hokkaido University (hp200130).
This work is supported in part by the Japanese Grant-in-Aid for Scientific Research (No. JP26247043, JP18H01216 and JP18H04484), and by MEXT as ``Priority Issue on Post-K computer" (Elucidation of the Fundamental Laws and Evolution of the Universe) and by Joint Institute for Computational Fundamental Science (JICFuS).

\bibliographystyle{JHEP}
\bibliography{lattice2023}
\end{document}